\documentclass[aps,prd,,twocolumn,showpacs,preprintnumbers]{revtex4}
\usepackage{graphicx}
\usepackage{epsfig}
\usepackage{color}
\usepackage{amsmath}
\usepackage{epsf,amssymb,amsfonts,amsmath}

\newcommand{\be}{\begin{eqnarray}}
\newcommand{\ee}{\end{eqnarray}}

\newcommand{\bea}{\begin{eqnarray}}
\newcommand{\eea}{\end{eqnarray}}

\definecolor{added}{rgb}{1,0.2,0}
\definecolor{changed}{rgb}{0.2,0,1}

\def \be {\begin{equation}}
\def \ee {\end{equation}}
\def \ba {\begin{array}}
\def \ea {\end{array}}
\def \bea{\begin{eqnarray}}
\def \eea{\end{eqnarray}}

\def \a {\alpha}
\def \b {\beta}
\def \g {\gamma}

\def \m {\mu}
\def \n {\nu}
\def \k {\kappa}

\def \s {\sigma}

\def \r {\rho}

\def \p {\partial}
\def \f {\frac}

\def \sr {\sqrt}

\def \and {{\textrm{and}}}

\begin{document}

\title{Locating the QCD critical end point through the peaked baryon number susceptibilities along the freeze-out line}
\author{Zhibin Li $^{a,b}$}
\author{Yidian Chen$^{a,b}$}
\author{Danning Li$^{c}$}
\author{Mei Huang$^{a,b,d}$}
\affiliation{$^{a}$ Institute of High Energy Physics, Chinese Academy of Sciences, Beijing 100049, China}
\affiliation{$^{b}$ School of Physics Sciences, University of Chinese Academy of Sciences, Beijing 100039, China}
\affiliation{$^{c}$ Department of Physics, Jinan University, Guangzhou 510632, China}
\affiliation{$^{d}$ Theoretical Physics Center for Science Facilities, Chinese Academy of Sciences, Beijing 100049, China}

\begin{abstract}
We investigate the baryon number susceptibilities up to fourth order along different freeze-out lines in a holographic QCD model with the critical end point (CEP), and we propose that the peaked baryon number susceptibilities along the freeze-out line can be used as a clean signature to locate the CEP in the QCD phase diagram. On the temperature and baryon chemical potential plane, the ratio of the baryon number susceptibilities (up to fourth order) forms a ridge along the phase boundary, and develops a sword shape mountain standing upright around the CEP in a narrow and oblate region. This feature is model independent and universal if the CEP exists. The measurement of  baryon number susceptibilities from heavy-ion collision experiment is along the freeze-out line. If the freeze-out line crosses the foot of the CEP mountain, then one can observe the peaked  baryon number susceptibilities along the freeze-out line, and the kurtosis of the baryon number distributions has the tallest magnitude. The data from the first phase of beam energy scan program at the Relativistic Heavy Ion Collisions indicates that a peak of the kurtosis of the baryon number distribution would show up at the collision energy around 5 GeV, which suggests that the freeze-out line crosses the foot of the CEP mountain and the summit of the CEP would be located nearby around the collision energy of 3 GeV.

\end{abstract}
\pacs{13.40.-f, 25.75.-q, 11.10.Wx }
\maketitle


Quantum Chromodynamics (QCD)  is the fundamental theory of the strong interactions. The QCD vacuum structure and its phase diagram under extreme conditions has been always one of the most attractive topics to understand the nonperturbative nature of the strong interactions. In the QCD vacuum, the chiral symmetry is spontaneously broken and color-charged quarks and gluons are confined. It is expected that at high temperature and/or baryon density, the system undergoes a phase transition from hadronic phase to chiral restored and deconfined quark-gluon phase. In the case of physical quark masses, QCD chiral models as well as lattice QCD predicted that the phase transition is of smooth crossover at small baryon chemical potentia and high temperature \cite{Fodor:2001au}.  Due to the sign problem, it is still quite challenging for lattice QCD simulation to work at finite baryon chemical potential. However, through symmetry classes analysis \cite{Pisarski:1983ms,Hatta:2002sj} and effective model predictions\cite{Schwarz:1999dj,Zhuang:2000ub,Chen:2014ufa,Chen:2015dra,Fan:2016ovc,Fan:2017kym,Fu:2010ay,Bowman:2008kc,Mao:2009aq,
Schaefer:2011ex,Schaefer:2012gy,Qin:2010nq,Luecker:2013oda,Fu:2016tey,Stephanov:2008qz,Stephanov:2011pb,Asakawa:2009aj,Athanasiou:2010kw}, it has been generally believed that QCD phase transition is of first order at high baryon chemical potential, and the end point of the first order phase transition line toward the crossover region is called the critical end point (CEP). For theoretical review of QCD phase diagram and the CEP, please refer to Refs. \cite{Stephanov:2004wx,Stephanov:2007fk} and references therein.

Different models such as Nambu--Jona-Lasinio (NJL) model, the Polyakov-loop improved NJL (PNJL) model, linear sigma model, quark-meson (QM) model, the Polyakov-loop improved QM model, and the Dyson-Schwinger equations (DSE) give various location of CEP, even the same model with different parameters have different location of the CEP \cite{Schwarz:1999dj,Zhuang:2000ub,Chen:2014ufa,Chen:2015dra,Fan:2016ovc,Fan:2017kym,Fu:2010ay,Bowman:2008kc,Mao:2009aq,
Schaefer:2011ex,Schaefer:2012gy,Qin:2010nq,Luecker:2013oda,Fu:2016tey}.  Therefore, to search for the existence of the CEP and to locate the CEP in the QCD phase diagram is one of the most central goals at the Relativistic Heavy Ion Collisions (RHIC), and it also sets a strong motivation for the future accelerator facilities at FAIR in Darmstadt and NICA in Dubna. In the first phase of beam energy scan program (BES-I) by the STAR and PHENIX experiments at RHIC runned from 2010 to 2014, the experimental measurements of the fluctuations of conserved quantities have been performed for Au+Au collisions at $\sqrt{s_{NN}}=7.7, 11.5, 14.5, 19.6, 27, 39, 62.4$ and $200 {\rm GeV}$.  The experimental measurements of cumulants of conserved quantities up to the fourth order  of net-proton, net-charge and net-kaon multiplicity distributions from BES-I \cite{Adamczyk:2013dal,Aggarwal:2010wy} are summarized in Ref.\cite{Luo-Xu}. One interesting observation is that the kurtosis of the baryon number distributions $\kappa \sigma^2$ in the most central Au+Au collisions shows a non-monotonic energy dependence behavior: It decreases from almost 1 at the colliding energy of $\sqrt{s_{NN}}=200 {\rm GeV}$ to 0.1 at $\sqrt{s_{NN}}=20{\rm GeV}$ then starts to increase quickly to 3.5 at $\sqrt{s_{NN}}=7{\rm GeV}$.

It is curious to know whether the non-monotonic behavior of  the kurtosis of the baryon number distributions is caused by the existence of the CEP.
Furthermore, before the running of the second phase of beam energy scan (BES-II) at RHIC in 2019-2020, it is urgent for theorists to offer a clean signature to identify the existence of the CEP, and even more to propose a method to locate the CEP. In this work, we are going to provide such an answer. It was shown in \cite{Stephanov:2011pb} that the quartic cumulant (or kurtosis) is universally negative when the critical point is approaching to the crossover side of the phase separation line. Previously, many interests have focused on the decreasing and then increasing behavior of the kurtosis of the baryon number distributions around the colliding energy $\sqrt{s_{NN}}=20{\rm GeV}$, which is caused by the sign changing of various cumulants around the CEP.  More sign changes for higher order susceptibilities have been recently discussed in \cite{Fan:2017kym}.

In this work, we will show that the cumulant (up to fourth order) of conserved number susceptibilities forms a ridge along the phase boundary on the $(T,\mu_B)$ plane, and if there exists the CEP, the universal landform feature for the conserved number susceptibility is that a high sharp hollow mountain or a sword shape mountain stands upright around the CEP. The measurement of  number susceptibilities from heavy-ion collision experiment is along the freeze-out line, which is below the phase boundary. If one is lucky enough that the freeze-out line can cross the foot of the CEP mountain, then one can observe the peaked  baryon number susceptibilities along the freeze-out line.

In recent decade, a new nonperturbative method has been developed based on anti-de Sitter/conformal field theory (AdS/CFT) correspondence and the conjecture of the gravity/gauge duality\cite{Maldacena:1997re,Gubser:1998bc,Witten:1998qj} to deal with strongly coupled QCD system. In the framework of holographic QCD (hQCD) model at finite baryon chemical potential, the CEP has been discussed in Refs. \cite{DeWolfe:2010he,DeWolfe:2011ts,Yang:2014bqa,Critelli:2017oub}.
The model we used in this work is a simple hQCD model with the CEP located at higher baryon chemical potential, and the 5D Einstein-Maxwell-dilaton holographic model is described by the action \cite{Yang:2014bqa}
\be
S=\f{1}{2\k_5^2} \int d^5 x \sr{-g}\left[R-\f{f\left(\phi\right)}{4}F_{\m\n}^2-\f{1}{2}(\p\phi)^2-V(\phi)\right],
\ee
and the ansatz of the metric is \cite{Yang:2014bqa}
\be
ds^2=\f{e^{2A(z)}}{z^2}[-g(z)dt^2+\f{1}{g(z)}dz^2+d\vec{x}^2].
\ee
With the regular boundary conditions at the horizon $z=z_H$ and the asymptotic $AdS_5$ condition at the boundary $z=0$ \cite{Yang:2014bqa}
\be
A_t(z_H)=g(z_H)=0,
\ee
\be
A(0)=-\sr{\f{1}{6}}\phi(0), \hspace{0.3cm} g(0)=1,
\ee
\be
A_t(0)=\m+\r z^2+\cdots.
\ee
where $\m$ and $\r$ are the chemical potential and density of quark respectively.
The warped factor and gauge kinetic function can be fixed as \cite{Yang:2014bqa}
\be
A(z)=-\frac{c}{3} z^2-b z^4,
\ee
\be
f(\phi(z))=e^{c z^2-A(z)}.
\ee
Then from the equation of motion we can calculate the quark density $\r$ and temperature $T$ as
\be
\r=\frac{c \m }{1-e^{c z_H^2}},
\ee
\begin{eqnarray}
& &T=\frac{z_H^3 e^{-3 A(z_H)}}{4 \pi  \int_0^{z_H} y^3 e^{-3 A(y)} \, dy}[1-\frac{2 c \mu ^2}{(1-e^{c z_H^2})^2}\times
  \nonumber \\
 & & (e^{c z_H^2} \int_0^{z_H} y^3 e^{-3 A(y)} \, dy-\int_0^{z_H} y^3 e^{c y^2-3 A(y)} \, dy)].
\end{eqnarray}
Here the parameters $b$ and $c$ are fixed from the meson spectrum and speed of sound \cite{Yang:2014bqa}
with $b=-6.25\times 10^{-4} {\rm GeV}^4$, and $c=0.227{\rm GeV}^2$. Note that here we have fixed $\k_5$ to $1$.

\begin{figure}[!thb]
\centerline{\includegraphics[width=8cm]{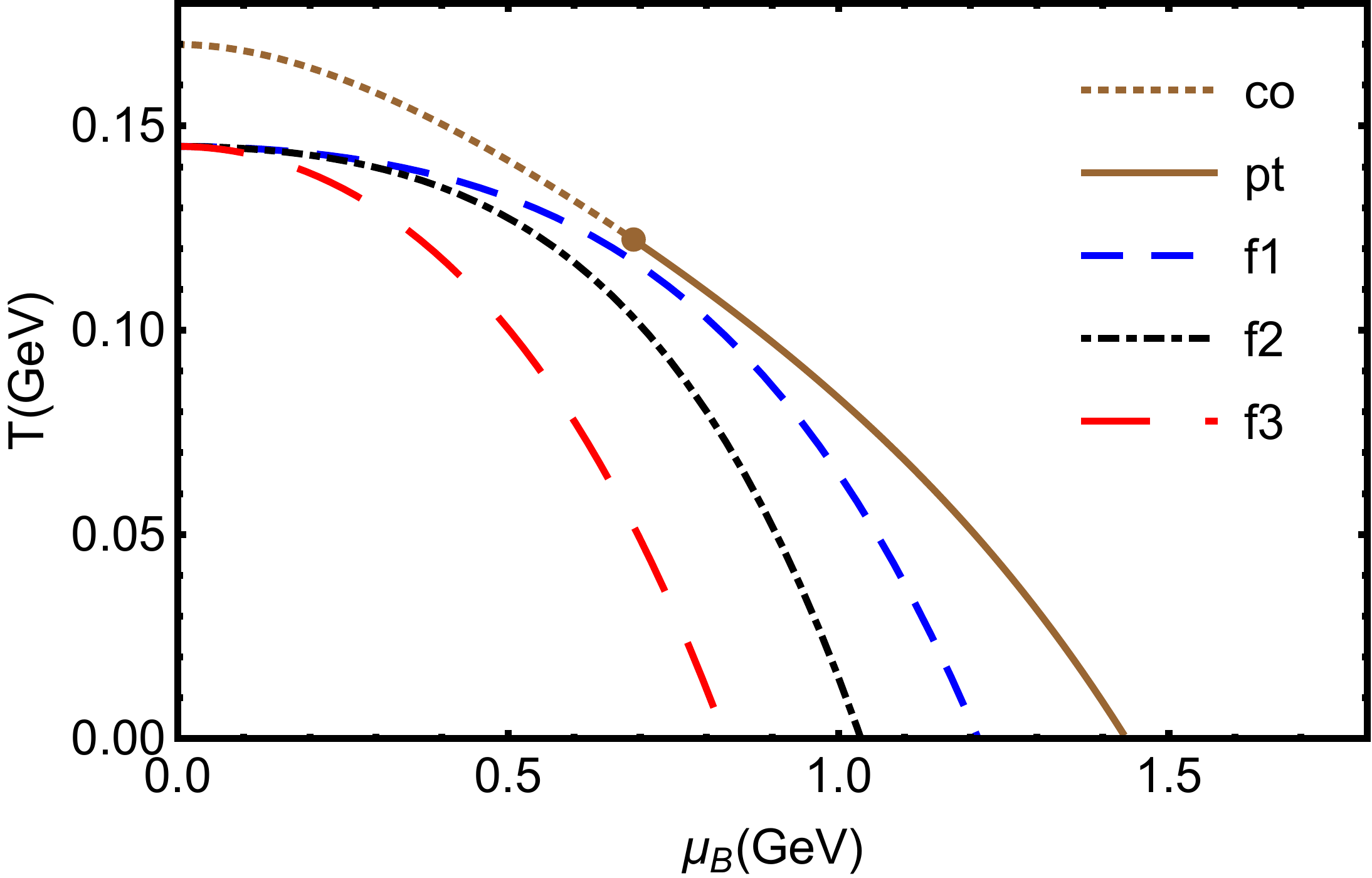}}
\caption{The $(T,\mu_B)$ phase diagram in the 5D hQCD model. The brown dotted line is for the crossover, and the brown solid line is for the first order phase transition, and the CEP is loated at  $(T^c,\mu_B^c)=(0.121{\rm GeV},0.693{\rm GeV})$. The blue dashed, black dashed-dotted, and the red long dashed lines represent three chemical freeze-out conditions defined in Eq.(\ref{Eq:fz}).}
\label{fig:phase}
\end{figure}

The entropy density $s$ can be calculated as \cite{DeWolfe:2011ts}
\be
s=2\pi \f{e^{3A(z_H)}}{z_H^3},
\ee
then by using the free energy \cite{Yang:2014bqa,DeWolfe:2010he}
\be
F=-\int [sdT+\r d\m],
\ee
one can determine the phase structure of this holographic QCD model as shown in Fig.\ref{fig:phase} with the CEP located at $(T^c,\mu_B^c)=(0.121{\rm GeV},0.693{\rm GeV})$, which is close to the CEP location given in \cite{Critelli:2017oub}. Using the parameterized relation between the collision
energy $\sqrt{s_{NN}}$ and $\m_B$ \cite{Kadeer:2005aq}
 \be
 \m_B(\sqrt{s_{NN}})=\f{1.30}{1+0.28\sqrt{s_{NN}}},
 \ee
the corresponding collision energy at CEP is $ \sqrt{s_{NN}}=2.71{\rm GeV}$.

The measurement from heavy-ion collision experiment is along the freeze-out line. We choose three polynomial fits for the chemical freeze-out line
\be
f_i: T(\m_B)=\a-\b_i\m_B^2-\g_i\m_B^4\hspace{0.5cm} i=1,2,3
\label{Eq:fz}
\ee
with $\a=0.145$, $\b_1=0.040$, $\b_2=0.060$, $\b_3=0.160$, $\g_1=\g_2=0.040$, $\g_3=0.074$.
The three chemical freeze-out lines $f_1$, $f_2$ and $f_3$ are shown in Fig.\ref{fig:phase} by
dashed, dashed-dotted and long dashed lines, respectively. The chemical freeze-out line $f_1$ is very close to the phase boundary, $f_3$
is away from the phase boundary, and $f_2$ is in between. In order to compare with experiment measurement, we choose the system freezes out
starting from $(T=145 {\rm MeV}, \mu_B=0)$.

The baryon number susceptibilities are defined as the derivative of the dimensionless pressure with respected to the reduced chemical potential \cite{Luo-Xu}
\be
\chi^B_n=\f{\p ^n[P/T^4]}{\p [\m_B /T]^n}
\ee
with the pressure $P=-F$ just the minus free energy \cite{DeWolfe:2011ts},
and the cumulants of baryon number distributions are given by
\be
C_n^B=VT^3\chi^B_n
\ee
Introducing the mean $M=C_1^B$, variance $ \s^2=C_2^B$, skewness $S=\f{C_3^B}{(\s^2)^{3/2}}$  and kurtosis $\k=\f{C_4^B}{(\s^2)^2}$,
one can have following relations between observable quantities and theoretical calculations
\begin{eqnarray}
\f{\s^2}{M}=\f{C_2^B}{C_1^B}=\f{\chi_2^B}{\chi_1^B},\,\, & &  S\s=\f{C_3^B}{C_2^B}=\f{\chi_3^B}{\chi_2^B}, \, \nonumber \\ \f{S\sigma^3}{M}=\f{C_3^B}{C_1^B}=\f{\chi_3^B}{\chi_1^B},\,\,& & \k \s^2=\f{C_4^B}{C_2^B}=\f{\chi_4^B}{\chi_2^B}.
\label{Eq:ratios}
\end{eqnarray}

\begin{figure*}[htbp]
  \centering
  \includegraphics[width=0.75\textwidth]{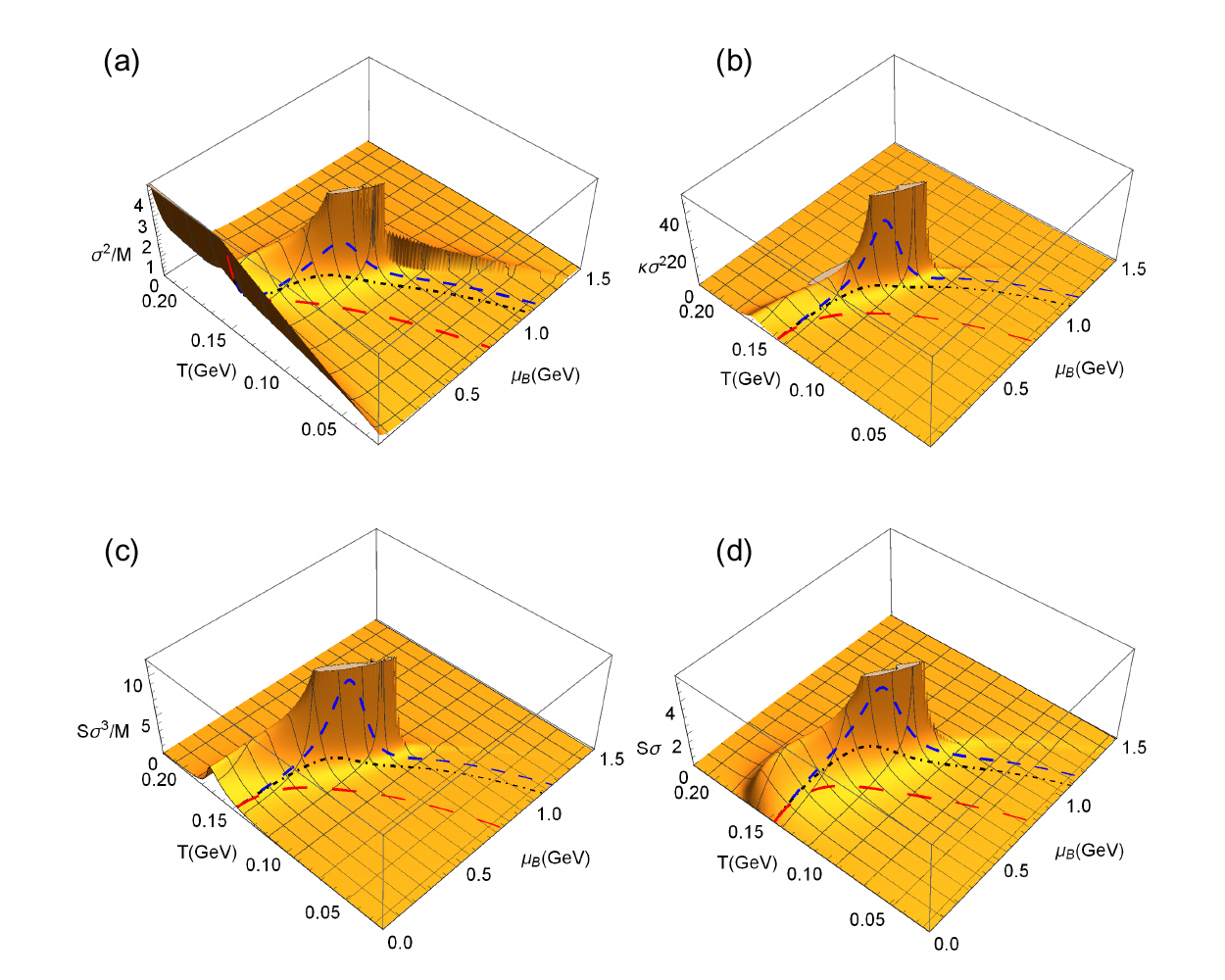}\\
  \caption{3D plot of ${\s^2}/{M}$, $S\s$,  ${S\sigma^3}/{M}$, $ \k \s^2$ as functions of the temperature $T$ and the baryon chemical potential $\mu_B$. The blue dashed, black dashed-dotted, and the red long dashed lines represent three chemical freeze-out conditions defined in Eq.(\ref{Eq:fz}).}
  \label{fig:3D_all}
\end{figure*}

Fig.\ref{fig:3D_all} shows the 3-dimension plot of ${\s^2}/{M}$, $S\s$,  ${S\sigma^3}/{M}$, $ \k \s^2$ as functions of the temperature $T$ and the baryon chemical potential $\mu_B$. It is observed that each ratio of the baryon number susceptibilities ${\s^2}/{M}$, $S\s$,  ${S\sigma^3}/{M}$, and $ \k \s^2$ forms an obvious ridge along the phase boundary, and it develops a high sharp hollow sword shape mountain standing upright around the CEP in a narrow oblate region. The CEP mountains are hollow because these ratios of the baryon number susceptibilities are negative inside this oblate region \cite{Stephanov:2011pb}. The profile along the phase boundary and the sword shape CEP mountain for baryon number susceptibilities (up to fourth order) is a universal feature if the CEP exists \cite{Chen:2014ufa,Chen:2015dra,Fan:2016ovc,Critelli:2017oub}, and this feature is independent of model used for analysis, and can extend to other conserved number susceptibilities.

The baryon number susceptibilities ${\s^2}/{M}$, $S\s$,  ${S\sigma^3}/{M}$, and $ \k \s^2$ along three chemical freeze-out lines $f_1$, $f_2$ and $f_3$ are also shown in Fig.\ref{fig:3D_all} by dashed, dashed-dotted and long dashed lines, respectively. If the freeze-out line is very close to the phase boundary as
$f_1$, it climbs up to the sword mountain of CEP a little bit, and the ratio of the baryon number susceptibilities ${\s^2}/{M}$, $S\s$,  ${S\sigma^3}/{M}$, or $ \k \s^2$ shows a high peak along the freeze-out line. If the freeze-out line is away from the phase boundary as $f_3$, it crosses the flat plane, and almost all the ratios of the baryon number susceptibilities of ${\s^2}/{M}$, $S\s$,  ${S\sigma^3}/{M}$, and $ \k \s^2$ show a monotonic decreasing behavior along the freeze-out line except that $S\s$. If the freeze-out line is not far away from the phase boundary as $f_2$, and it can cross the foot of the sword mountain of CEP, the ratio of the baryon number susceptibilities ${\s^2}/{M}$, $S\s$,  ${S\sigma^3}/{M}$, or $ \k \s^2$ shows an obvious peak along the freeze-out line. The closer the freeze-out line to the phase boundary, the higher the peak of the number susceptibilities. Among the four ratios of the baryon number susceptibilities ${\s^2}/{M}$, $S\s$,  ${S\sigma^3}/{M}$, or $ \k \s^2$, it is found that the peak of the kurtosis $ \k \s^2$ has the tallest magnitude for the same freeze-out condition.

\begin{figure*}[htbp]
  \centering
  \includegraphics[width=0.75\textwidth]{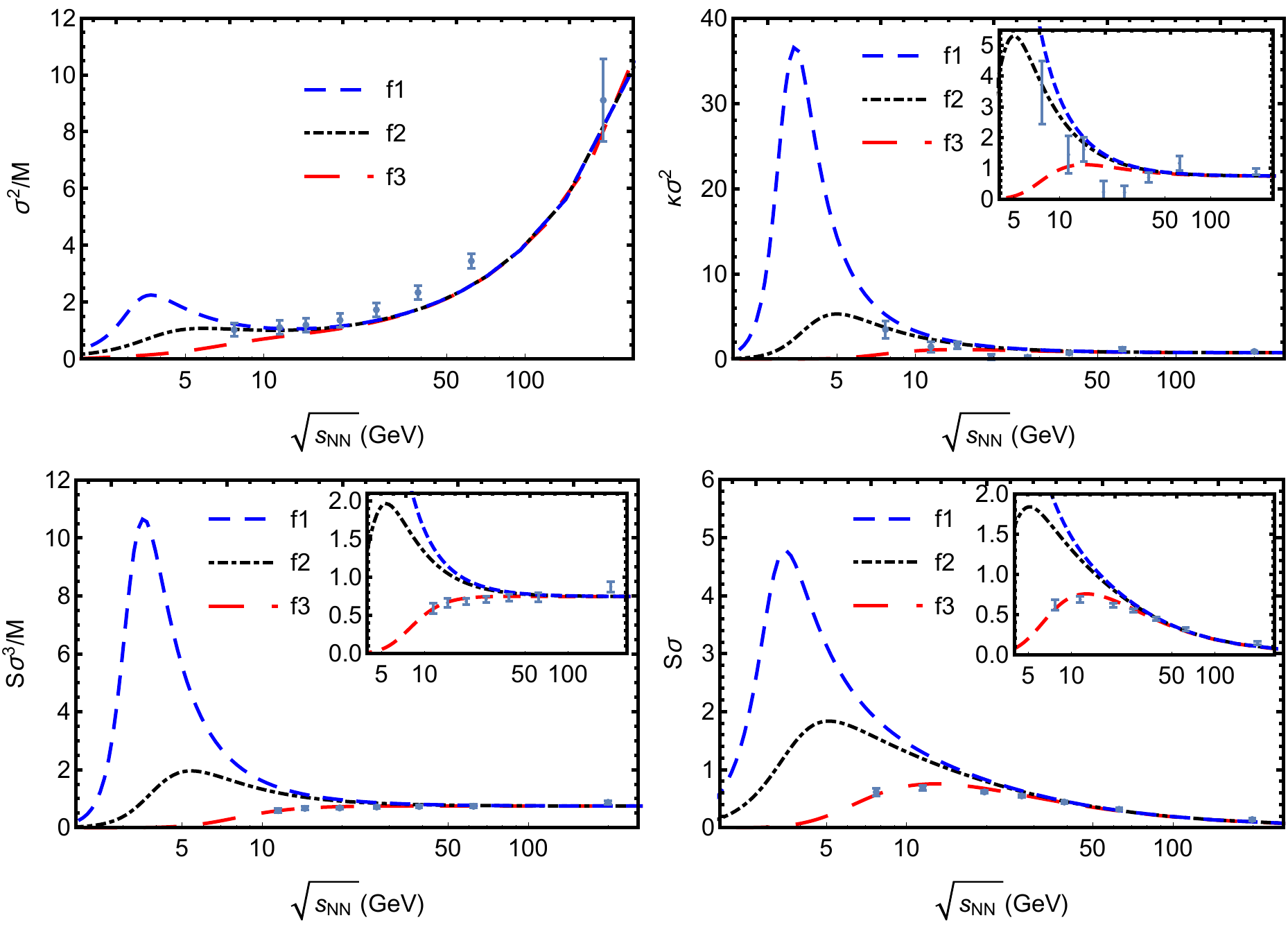}\\
  \caption{${\s^2}/{M}$, $S\s$,  ${S\sigma^3}/{M}$, $ \k \s^2$ along freeze-out lines as a function of the collision energy $\sqrt{s_{NN}}$ and comparing with experiment measurement. The blue dashed, black dashed-dotted, and the red long dashed lines represent three chemical freeze-out conditions defined in Eq.(\ref{Eq:fz}).}
  \label{fig:2D_all}
\end{figure*}

In Fig.\ref{fig:2D_all}, we compare our model results with experiment measurement of the baryon number susceptibilities ${\s^2}/{M}$, $S\s$,  ${S\sigma^3}/{M}$, and $ \k \s^2$ along three freeze-out lines $f_1,f_2$ and $f_3$ as a function of the collision energy $\sqrt{s_{NN}}$. It is observed that above the collision energy of $\sqrt{s_{NN}}=20 {\rm GeV}$, the model results along all three freeze-out lines are in agreement with experiment results very well. The experiment measurement of even cumulants ${\s^2}/{M}$ ($C_2^B/C_1^B$) and $ \k \s^2$ ($C_4^B/C_2^B$ ) follows the freeze-out line $f_2$, while experiment measurement of the odd cumulants ${S\sigma^3}/{M}$ ($C_3^B/C_1^B$ ) and $S\s$($C_3^B/C_2^B$) goes along the freeze-out line $f_3$. It might be due to the reason that different bases of experiment data are chosen for experimental analysis \cite{discussion}.

Because higher cumulant has higher magnitude around CEP, therefore we only focus on the kurtosis measurement $ \k \s^2$. The measurement of $ \k \s^2$ follows the freeze-out line $f_2$, the current data indicates that it would show up a peak around the collision energy $\sqrt{s_{NN}}=5 {\rm GeV}$.  As we have explained above that because the freeze-out line $f_2$ is close to the phase boundary, and crosses the foot of the CEP mountain, one can observe an obvious peak of the baryon number susceptibilities along the freeze-out line. The peaked kurtosis $ \k \s^2$ along the freeze-out line gives an evident signature of the existence of the CEP, and we can estimate that the CEP is located nearby around the collision energy $\sqrt{s_{NN}}=3 {\rm GeV}$.

In summary, we have explained that why the peaked baryon number susceptibilities especially the kurtosis $ \k \s^2$ along the freeze-out line can be used as an evident signature for the existence of the CEP and the peak position can also be used to locate the rough region of the CEP in the QCD phase diagram. If the CEP exists, there is a universal feature for the 3D profile of baryon number susceptibilities (up to fourth order) that it forms a ridge along the phase boundary and develops a sword shape mountain standing upright around the CEP in a narrow oblate region. This feature is model independent and can be extend to other conserved number susceptibilities. It should be emphasized that the sword shape CEP mountain stands on the $(T,\mu_B)$ plane in a very narrow and oblate region along the phase boundary, the foot area is also quite narrow and oblate. Therefore, if we can observe the peaked kurtosis $ \k \s^2$ along the freeze-out line, this means that the real freeze-out line can cross the narrow foot of the CEP mountain. That would be very lucky for experimentalists!  Our analysis estimates that the peak of the  kurtosis $ \k \s^2$ would show up around the collision energy $\sqrt{s_{NN}}=5 {\rm GeV}$, and the CEP would be located around  the collision energy $\sqrt{s_{NN}}=3 {\rm GeV}$. Finally, we would like to add one comment on the finite size effect, which might broaden a little bit the CEP region, and it would give more chances for experimentalists to find the CEP. 

\begin{acknowledgments}
We thank X.F.Luo and I.Shovkovy for valuable discussions. This work is supported by the NSFC under Grant No. 11275213, and 11261130311(CRC 110 by DFG and NSFC), CAS key project KJCX2-EW-N01, and Youth Innovation Promotion Association of CAS.
\end{acknowledgments}

\end{document}